# Nucleon Electromagnetic Form Factors[*]

Kees de Jager

*Jefferson Laboratory, Newport News, VA 23606, USA*

**Abstract.** A review of data on the nucleon electromagnetic form factors in the space-like region is presented. Recent results from experiments using polarized beams and polarized targets or nucleon recoil polarimeters have yielded a significant improvement on the precision of the data obtained with the traditional Rosenbluth separation. Future plans for extended measurements are outlined.

## INTRODUCTION

The nucleon electromagnetic form factors (EMFF) are of fundamental importance for an understanding of their internal structure. These EMFF, which in the Breit frame can be simply related to the spatial distribution of the nucleon charge and magnetization densities, are measured through elastic electron-nucleon scattering.

In Plane Wave Born Approximation (PWBA) the cross section can be expressed in terms of the so-called Sachs form factors $G_E$ and $G_M$

$$\frac{d\sigma}{d\Omega} = \frac{d\sigma}{d\Omega_M} f_{rec}^{-1} \left\{ \frac{G_E^2 + \tau G_M^2}{1+\tau} + 2\tau G_M^2 \tan^2(\theta_e / 2) \right\} \tag{1}$$

$$f_{rec} = 1 + \frac{2E_e}{m_N} \sin^2(\theta_e / 2) \qquad \tau = \frac{Q^2}{4m_N^2}$$

where $Q$ is the four-momentum transfer, $\sigma_M$ the Mott cross section for scattering off a point-like particle, $m_N$ the mass of the nucleon, $\theta_e$ the electron scattering angle and $E_e$ the electron energy. This equation shows that $G_E$ and $G_M$ can be determined separately by measuring at fixed $Q^2$ over a range of $(\theta_e, E_e)$ combinations. This procedure is called the Rosenbluth separation.[1]

---

The Sachs form factors can be identified with the Fourier transform of the nucleon charge and magnetization density distributions, such that the slope at $Q^2 \to 0$ of the EMFF is related to the charge and magnetization radius. There has been considerable debate over the interpretation of the neutron charge radius.[2,3] The charge radius can be split into two components, one of which (the so-called Foldy[4] term) is related to the magnetic moment, not to the rest-frame charge distribution. However, Isgur[5] and Bawin and Coon[6] have shown that this Foldy term is exactly canceled by a contribution from the Dirac form factor, so that the charge radius is indeed determined solely by the rest-frame charge distribution.

Up until the beginning of the previous decade all available proton EMFF data had been collected using the Rosenbluth separation. This experimental procedure requires an accurate knowledge of the electron energy and the total luminosity. In addition, since the contribution to the elastic cross section from the magnetic form factor is weighted with $Q^2$, data on $G_E^p$ suffer from increasing systematic uncertainties at higher $Q^2$-values. Data for the neutron resulted mainly from quasi-elastic scattering off the deuteron, because a free neutron target is not available in nature. This additional constraint caused large uncertainties, especially on the data for $G_E^n$.

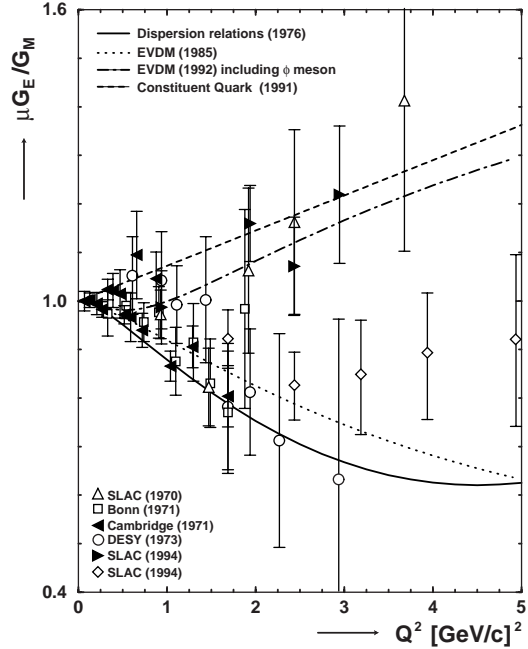

**FIGURE 1.** The ratio $\mu_p G_E^p / G_M^p$ as a function of $Q^2$, determined with the Rosenbluth separation technique. Data symbols are explained in ref. 48. Theory: full[9], dotted[11], dashed[16] and dot-dashed.[12]

These restrictions are clearly presented in the review paper by Bosted et al.[7] The then available world data set was compared to the so-called dipole parametrization, which corresponds to exponentially decreasing radial charge and magnetization densities:

$$G_E^p = G_D \qquad\qquad G_E^n = 0$$
$$G_M^p = \mu_p G_D \qquad\qquad G_M^n = \mu_n G_D \qquad\qquad\qquad (2)$$
$$G_D = \left(1 + \frac{Q^2}{Q_o^2}\right)^{-2} \qquad \text{with } Q \text{ in } GeV/c \text{ and } Q_0 = 0.84 \; GeV/c$$

Accurate data were available for $G_M^p$ up to $Q^2$-values of over 20 (GeV/c)$^2$, whereas for $G_E^n$ no significant deviation from zero was measured.[8] For all four EMFF the available data agreed with the dipole parametrization to within 20 %. However, the limitation of the Rosenbluth separation is evident from fig. 1, which shows all available data on $G_E^p$. Different data sets deviate from each other by up to 50 % at higher $Q^2$-values, way beyond the already sizeable estimate for the experimental uncertainty.

## THEORY

A frequently used framework[9] to describe the EMFF is that of Vector Meson Dominance (VMD), in which one assumes that the virtual photon – after having become a quark-antiquark pair - couples to the nucleon as a vector meson. The EMFF can then be expressed in terms of coupling strengths between the virtual photon and the vector meson and between the vector meson and the nucleon, summing over all possible vector mesons. In some cases additional terms are included to account for the effect of unknown or lesser known mesons.

A common restriction of the VMD models is that they do not predict a correct behaviour of the EMFF at high $Q^2$-values. The quark-dimensional scaling framework[10] predicts that only valence quark states contribute at sufficiently high $Q^2$-values. Under these conditions the EMFF $Q^2$-dependence is determined simply by the number of gluon propagators, causing the Dirac and Pauli form factors to be proportional to $Q^{-4}$ and $Q^{-6}$, respectively, whereas any VMD-model will predict a $Q^{-2}$ behaviour at large $Q^2$-values. Gari and Krümpelmann have constructed a hybrid (EVMD) model which combines the low $Q^2$-behaviour of the VMD model with the asymptotic behaviour predicted by pQCD. In their first paper[11] they consider only coupling to the $\rho$ and $\omega$ mesons, whereas later[12] the $\phi$ meson was also included.

VMD models form a subset of models using dispersion relations, which relate form factors to spectral functions. These spectral functions can also be thought of as a superposition of vector meson poles, but include contributions from n-particle production continua. This framework allows then a model-independent fit[13] to all available EMFF data in the space- and the time-like region.

Many attempts have been made to enlarge the domain of applicability of pQCD calculations to moderate $Q^2$-values. Kroll et al.[15] have generalized the hard-scattering scheme by assuming nucleons to consist of quarks and diquarks. The diquarks are used to approximate the effects of correlations in the nucleon wave function. This model is equivalent to the hard-scattering formalism of pQCD in the limit $Q^2 \rightarrow \infty$. Chung and Coester[16] have developed a relativistic constituent quark model with effective quark masses and a confinement scale as free parameters.

Lu et al.[17] have recently expanded the cloudy bag model, whereby the nucleon is described as a bag containing three quarks, but including an elementary pion field coupled to them, in such a way that chiral symmetry is restored. Finally, recent developments[18] within the Skewed Parton Distribution formalism indicate a relation between the EMFF behaviour at larger $Q^2$-values and the nucleon spin.

## NUCLEON FORM FACTORS

Over 20 years ago Akhiezer and Rekalo[19] showed that the accuracy of EMFF measurements could be increased significantly by scattering polarized electrons off a polarized target (or by equivalently measuring the polarization of the recoiling nucleon). In the early nineties a series of measurements[20-25] at the MIT-Bates facility showed the feasibility of that measurement principle.

### Neutron Magnetic Form Factor

Significant progress has been made in measurements of $G_M^n$ at low $Q^2$-values by measuring the ratio of quasi-elastic neutron and proton knock-out from a deuterium target. This method is practically insensitive to nuclear binding effects and to fluctuations in the luminosity and detector acceptance. The basic set-up used in all such measurements was very similar: the electron was detected in a magnetic spectrometer with coincident neutron/proton detection in a large scintillator array. The main technical difficulty in such a ratio measurement is the absolute determination of the neutron detection efficiency. For the measurements at Bates[25] and ELSA[26] the efficiency was measured in situ using the $D(\gamma,p)n$ or $p(\gamma,\pi^+)$ reaction with a bremsstrahlung radiator up stream of the experimental target. The hadron detectors used in the experiments at NIKHEF[27] and Mainz[28] were calibrated at the PSI neutron beam using the kinematically complete $p(n,p)n$ reaction.

Figure 2 shows the results of those four experiments. The Mainz $G_M^n$ data are 8-10 % lower than the ELSA ones, despite the quoted uncertainty of appr. 2 %. This discrepancy would require a 16-20% error in the detector efficiency. The contribution from electroproduction in the ELSA set-up, caused by the electron contamination in the bremsstrahlung beam, which could result in a loss of events due to the three-body kinematics in electroproduction, has been extensively investigated.[29] Thus far, the detection inefficiency due to electroproduction has been established at less than 5 %, clearly much smaller than required to explain the discrepancy in the data.

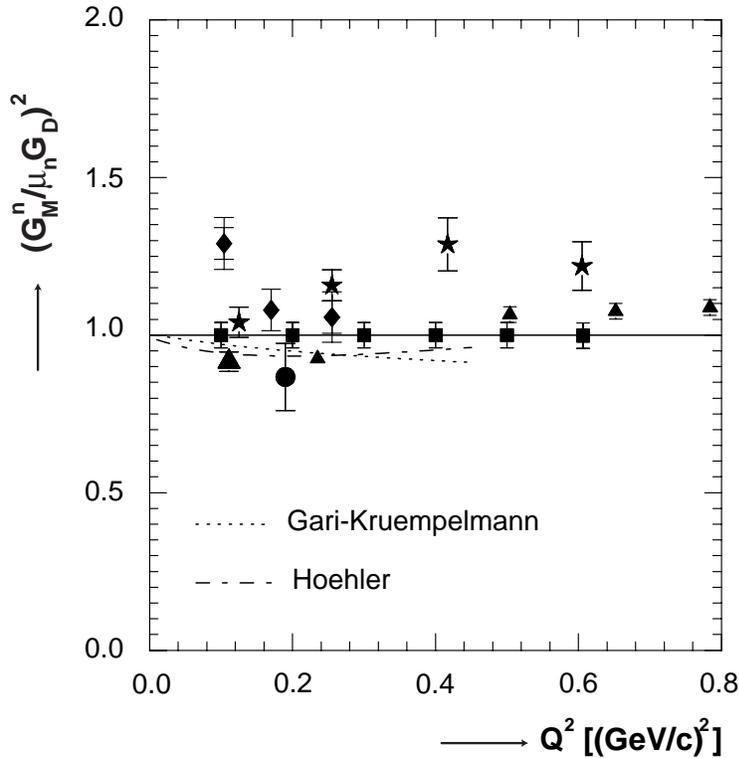

**FIGURE 2.** The square of the ratio of $G_M^n$ to $\mu_n G_D$ as a function of $Q^2$, compared to predictions by Gari and Krümpelmann[11] and Höhler[9]. The expected precision of JLab experiment E95-001[30] is indicated by the solid squares. Data: diamonds[25], stars[26], circle[24], large triangle[27], triangles.[28]

Recently, inclusive quasi-elastic scattering of polarized electrons off a polarized $^3$He target was measured[30] in Hall A at JLab in a $Q^2$-range from 0.1 to 0.6 (GeV/c)$^2$. This experiment will provide an independent accurate measurement of $G_M^n$ in a $Q^2$-range overlapping with that of the ELSA and Mainz data. Measurements of $G_M^n$ at $Q^2$- values up to 5 (GeV/c)$^2$ are expected in the near future from a JLab experiment that will measure the neutron/proton quasi-elastic cross-section ratio using the CLAS detector.[31]

## Neutron Electric Form Factor

Since a free neutron target is not available, one has to use neutrons bound in nuclei to study the neutron EMFF. The most precise data on $G_E^n$ prior to any spin-dependent experiment were obtained from the elastic electron-deuteron scattering experiment by Platchkov et al.[32] The deuteron elastic form factor contains a term of the form $G_E^n G_E^p$. However, in order to extract $G_E^n$ from the data, one has to calculate the deuteron wave function, which requires a choice of the nucleon-nucleon potential. Figure 3 shows the $G_E^n$ values extracted from the Platchkov data with the Paris potential, while the grey band indicates the range of $G_E^n$ values extracted with the Nijmegen, AV14 and RSC potentials. Clearly, the choice of NN-potentials results in a systematic uncertainty of appr. 50 % in $G_E^n$. One should realize that all modern NN-potentials yield consistent results for a large variety of two- and three-nucleon observables. Thus, one might expect that a reevaluation of the Platchkov data using modern high-precision NN-potentials and a consistent treatment of exchange currents will yield a reduced potential dependence.

Significant advances have been made in the last decade in the development of electron beams with high polarization and intensity and of reliable polarized targets. This progress has been used in a series of new spin-dependent measurements of $G_E^n$, which utilizes the fact that the ratio of the beam-target asymmetry with the target polarization perpendicular and parallel to the momentum transfer is directly proportional to the ratio of the electric and magnetic form factors:

$$\frac{G_E^n}{G_M^n} = \frac{A_\perp}{A_{//}} \sqrt{\tau + \tau(1+\tau)\tan^2(\theta_e / 2)} \tag{3}$$

A similar relation can be derived for the reaction $^2H(\vec{e},e'\,\vec{n})$ when one measures the polarization of the recoiling neutron directly and after having precessed the neutron spin over 90° with a dipole magnet. Figure 3 shows the results of the pioneering experiments of that technique, performed at Bates, using the reactions $^2H(\vec{e},e'\,\vec{n})$[20] and $^3\vec{He}(\vec{e},e')$[21-23] and at Mainz, with the $^3\vec{He}(\vec{e},e'n)$ reaction.[33] These results have not been corrected for rescattering or nuclear medium effects.

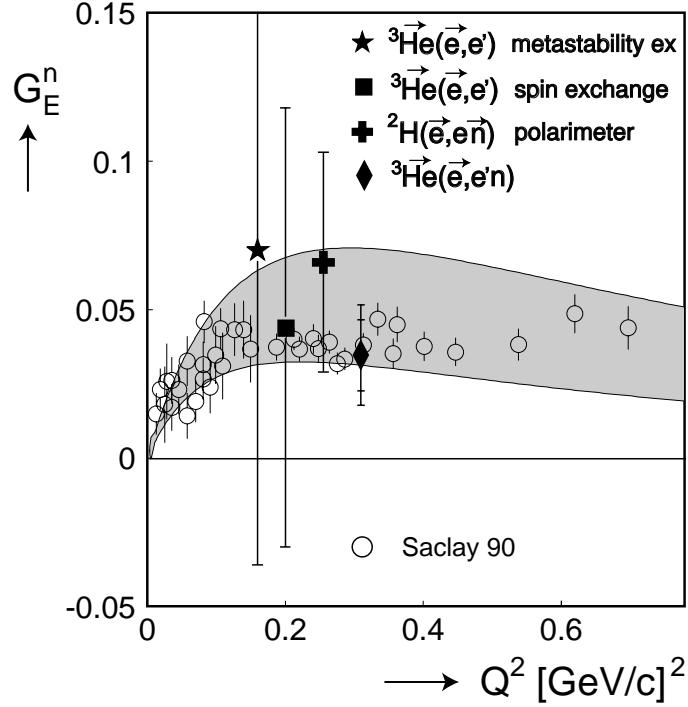

**FIGURE 3.** Older (star[21,22], square[23], cross[20] and diamond[33]) results for $G_E^n$ as a function of $Q^2$. The open circles depict the results of Platchkov et al.[32] for the Paris potential, the shaded area the systematic uncertainty due to the choice of NN-potential.

Figure 4 shows the most recent results, obtained through the reaction channels $^2\vec{H}(\vec{e},e'\,n)$[34], $^2H(\vec{e},e'\,\vec{n})$[35,36] and $^3\vec{He}(\vec{e},e'\,n)$.[37,38] At low $Q^2$-values corrections for nuclear medium and rescattering effects can be sizeable: 65 % for deuterium at 0.15 $(GeV/c)^2$ and 50 % for $^3$He[38] at 0.35 $(GeV/c)^2$. These corrections are expected to decrease significantly with increasing $Q$, although no reliable results are at present available for $^3$He above 0.5 $(GeV/c)^2$. Thus, there are now data from a variety of reaction channels available in a $Q^2$-range up to 0.6 $(GeV/c)^2$ with an overall accuracy of appr. 20 %, which are in mutual agreement. However, neither the VMD[11] nor the dispersion relation[14] calculations agree with the data. Only the Galster parametrization[40] which uses a modified version of the dipole form factor, is able to describe the data adequately. A more detailed discussion of these recent results is given by Schmieden.[41] Also shown in fig. 4 are the results expected in the near future, from the $^3\vec{He}(\vec{e},e'\,n)$ channel at NIKHEF[42] and from the $^2\vec{H}(\vec{e},e'\,n)$[43] and $^2H(\vec{e},e'\,\vec{n})$[44] channels at JLab. Finally, in fig. 5 are shown the results expected with the BLAST detector[45] with both the $^2\vec{H}(\vec{e},e'\,n)$ and the $^3\vec{He}(\vec{e},e'\,n)$ reaction channels.

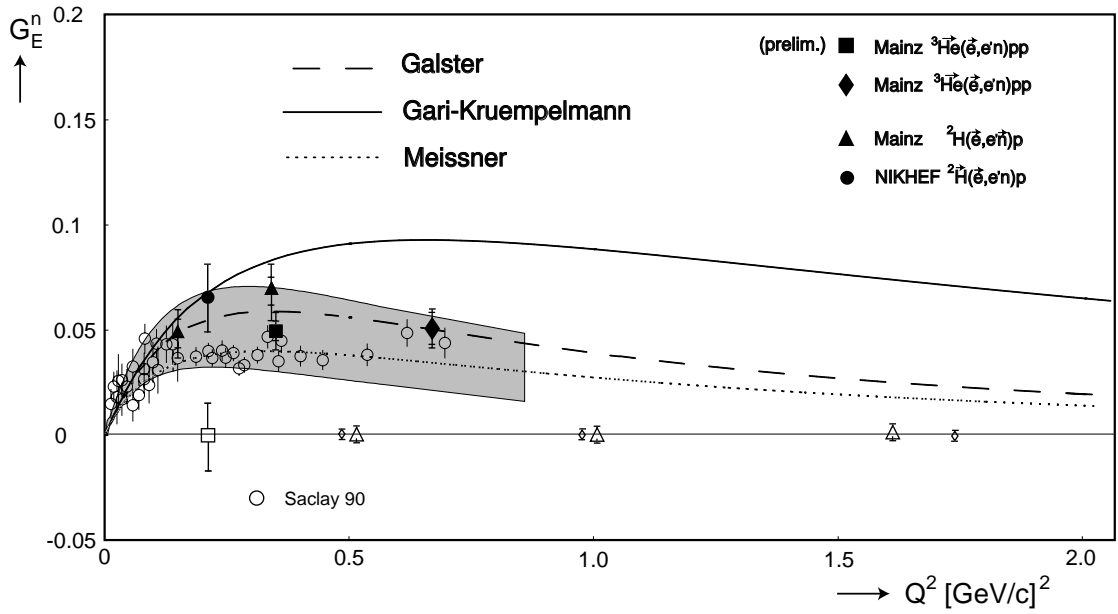

**FIGURE 4.** Recent (circle[34], triangle[35,36], square[37,39] and diamond[38]) and future (open square[42], open diamonds[44] and open triangles[43]) results for $G_E^n$ as a function of $Q^2$, compared to three theoretical calculations (full[11], dashed[14] and dotted[40]).

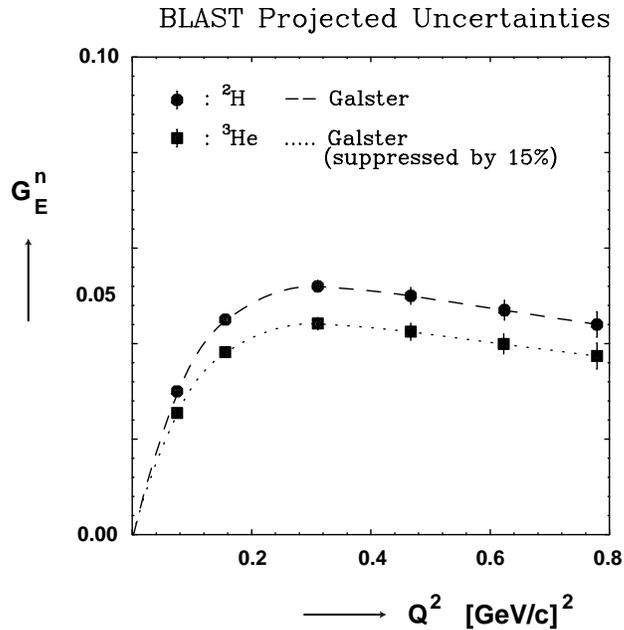

**FIGURE 5.** Predicted accuracy of $G_E^n$ data to be obtained with the BLAST detector.

## Proton Electric Form Factor

Arnold et al.[46] have shown that the systematic error in a measurement of $G_E^p$, inherent to the Rosenbluth separation, can be significantly reduced by scattering longitudinally polarized electrons off a hydrogen target and measuring the ratio of the transverse to longitudinal polarization of the recoiling proton.

$$\frac{G_E^p}{G_M^p} = -\frac{P_t}{P_l}\frac{\left(E_e + E_{e'}\right)}{2m_p}\tan\left(\theta_e / 2\right) \tag{4}$$

This ratio of the two polarization components can be measured in a focal plane polarimeter, while neither the beam polarization nor the polarimeter analyzing power need be known. This method was first used by Milbrath et al.[47] at MIT-Bates to measure the ratio $G_E^p / G_M^p$ at low $Q^2$.

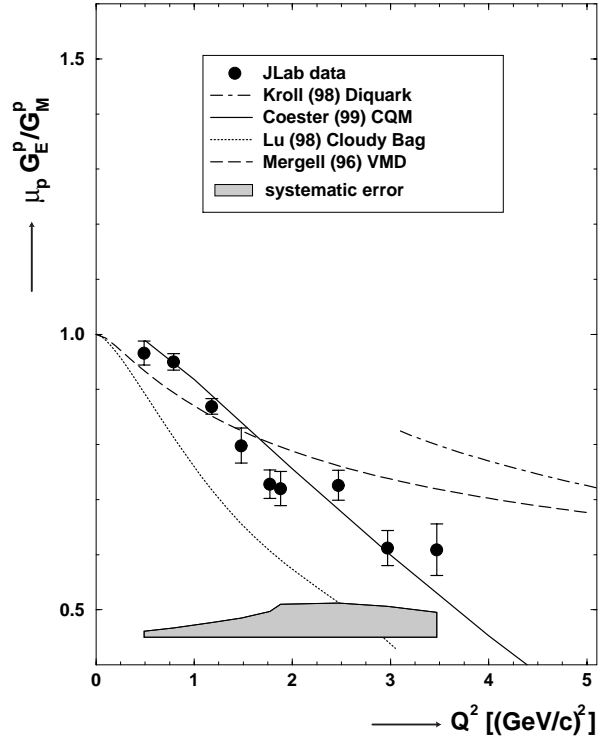

FIGURE 6. The JLab data[48] for the ratio $\mu_p G_E^p / G_M^p$ as a function of $Q^2$, compared to recent theoretical predictions (full[16], dotted[17], dashed[13] and dot-dashed[15]). The shaded area denotes the size of the systematic error in the data.

Recently a similar experiment[48] was performed in Hall A at JLab. Longitudinally polarized electrons with energies between 0.9 and 4.1 GeV were scattered in a 15 cm long liquid hydrogen target. For the four highest $Q^2$-values the beam conditions were 39 % polarization at currents up to 115 μA, while at the lower $Q^2$-values a 60 % polarization was obtained at currents up to 15 μA. Elastic $ep$ events were selected by detecting electrons and protons in coincidence in the two identical HRS spectrometers. The polarization of the recoiling proton was determined with a Focal Plane

Polarimeter (FPP) in the hadron HRS, consisting of two pairs of straw chambers with a carbon analyzer in between. Instrumental asymmetries are cancelled by taking the difference of the azimuthal distributions of the protons scattered in the analyzer for positive and negative beam helicity. A Fourier analysis of this difference then yields the transverse and normal polarization components at the FPP. The data were analyzed in bins of each of the target coordinates. No dependence on any of these variables was observed.

The results for the ratio $G_E^p / G_M^p$ are shown in fig. 6. The most striking feature of the data is the sharp decline as $Q^2$ increases. Since it is known that $G_M^p$ closely follows the dipole parametrization, it follows that $G_E^p$ falls more rapidly with $Q^2$ than the dipole form factor $G_D$. A comparison with fig. 1 confirms the expected improvement in accuracy of such a spin-dependent measurement over the Rosenbluth separation. None of the theoretical models shown in fig. 6 is able to adequately describe the new data. An extension[49] of this experiment to a $Q^2$-value of 5.6 $(\text{GeV/c})^2$ has been scheduled for the fall of 2000.

## CONCLUSIONS

Recent advances in polarized electron sources, polarized nucleon targets and nucleon recoil polarimeters have made it possible to accurately measure the spin-dependent elastic electron-nucleon cross section. New data on nucleon electromagnetic form factors with an unprecedented precision have (and will continue to) become available in an ever increasing $Q^2$-domain. These data will form tight constraints on models of nucleon structure and will hopefully incite new theoretical efforts. In addition they will significantly improve the accuracy of the extraction of strange form factors from parity-violating experiments.[50]

## ACKNOWLEDGMENTS

The author expresses his gratitude to Ulf Meissner and Mark Jones for fruitful discussions and for receiving their results prior to publication. This work was supported in part by the U.S. Department of Energy.